# Effective Steganography Detection Based On Data Compression

*Nechta I.*

This article describes novel text steganalysis method. The archiver "Bzip2" used for detection stegotext generated by Texto stegosystem. Experiments show that proposed approach gets better performance than typical existing methods. The detection accuracy exceeds 99.98% for text segments with size 400 bytes.

*Key words*: Steganography, steganalysis, linguistic stegosystem, text steganography.

(**The main idea**: The compression used for stegotext detection. It is known that an embedding message breaks statistical structure of the container, increasing its entropy. Consequently, the full container will compress worse than empty. Let us consider the example: A, B - empty and full container respectively.

Table 1: Container size before and after compression container..

| container | size before compr. | size after compr |
|---|---|---|
| A | 500 | 320 |
| B | 500 | 300 |

Add content of suspected container C into A and B. Compress them. Compare added content sizes
before and after compression.

Table 2: Content size of C container before compr. after compr.

| container | size before compr. | size after compr |
|---|---|---|
| $C_A$ | 50 | 45 |
| $C_B$ | 50 | 20 |

It could be asserted that container C is statistically depend with B, which ensures good compression. This principle used in attack on Texto. The detection accuracy exceeds 99.98% when text size is larger than 400 bytes.) See this article below:

**E-mail**: www@inbox.ru

# Эффективный метод стегоанализа базирующийся на сжатии данных

## И. В. Нечта

Предлагается новый метод обнаружения скрытых сообщений в лингвистической стегосистеме. В отличие от предыдущих методов, данный метод обладает высокой степенью обнаружения при малых объемах входных данных и не требует много времени работы.

*Ключевые слова*: Стеганография, стегоанализ, лингвистические стегосистемы, текстовая стеганография.

## 1. Введение

Задача стеганографии состоит в организации обмена секретными сообщениями между участниками так, чтобы сам факт обмена оставался скрытым от стороннего наблюдателя. Для этого в обычное сообщение, называемое *контейнером,* встраивают секретное сообщение с помощью специальных алгоритмов. Контейнер подбирается таким образом, чтобы содержание и сам факт его передачи не вызывал ни каких подозрений у стороннего наблюдателя. На сегодняшний день в сети Интернет передается большое число файлов различных типов, например, цифровые фотографии, видео, текст или музыка. Следовательно, такие файлы могут выступать в качестве контейнера. В данной статье речь пойдет об одном из направлений стеганографии, которое использует в качестве контейнера текстовые файлы.

Существующие методы «встраивания» секретных сообщений в текстовые данные, можно разделить на три группы:

*Синтаксические методы*. К таким методам можно отнести, например предложенный в работе [1], использующий дополнительные пробелы между словами. Один пробел соответствует нулю, два – единице. Данный метод может широко применяться в файлах формата HTML (интернет страниц), поскольку наличие пробелов никак не влияет на отображение страницы. Недостатком можно считать лёгкую обнаруживаемость, т.к. обычно при написании текста дополнительные пробелы не используются. Существует возможность использовать специальные символы вместо пробелов, не отображающиеся в часто используемых текстовых редакторах.

Еще один метод, предложенный в работе [1], использует синтаксические ошибки при написании слов, например:

"This is the end"
"This iz the end"

Во втором варианте допущена опечатка. Наличие опечатки в определенных словах (в частности "iz") означает, что бит передаваемой информации равен нулю, а отсутствие – единице. Таким образом, происходит передача информации в тексте. Данный метод не является легко обнаруживаемым, т.к. в обычном тексте ошибки также могут встречаться.

*Семантические методы*. К этой группе относят **Tyrannosaurus Lex**, опубликованный в работе [2], использующий замену слов в предложении на их синонимы, например:

|  | **excellent** | **city** |
|---|---|---|
| Tobolsk is a | (0) decent | (0) metropolis |
|  | (1) fine | (1) little town |

В зависимости от выбранного синонима кодируется передаваемое сообщение. Предложение "Tobolsk is a decent little town" содержит стегосообщение − "01". Данный метод требует наличия большого словаря синонимов. К недостатку таких методов относят возможное нарушение стиля написания текста. Например, (0) *. . . and make it still better, and say nothing of the bad–belongs to you alone.* (1) *. . . and make it still better, and say nada of the bad–belongs to you alone.* Слово "*nada*" является не типичным для использования некоторыми авторами, в частности, Jane Austen.

Также существует метод, опубликованный в работе [3], преобразующий обычный текст в стеготекст путем перефразирования предложений. Например, (0) *The caller identified the bomber as Yussef Attala, 20, from the Balata refugee camp near Nablus.* (1) *The caller named the bomber as 20-year old Yussef Attala from the Balata refugee camp near Nablus.* Данный метод обладает высокой степенью скрытности.

*Методы, генерирующие текст, подобный естественному.* Рассмотрим метод, предложенный в работе [4], использующий контекстно-свободные грамматики для генерации естественно подобного текста. Правила:

$$S \to A\ B\ C$$
$$A \to She\ (0)\ |\ He\ (1)$$
$$B \to likes\ (0)\ |\ hates\ (1)$$
$$C \to milk\ (0)\ |\ apples\ (1)$$

В зависимости от бита передаваемого сообщения выбираем правило раскрытия нетерминального символа. Соответственно если необходимо закодировать сообщение "101" получится: "He likes apples". На сегодняшний день наиболее популярными стегосистемами являются Nicetext[5], Texto[6] и Markov-Chain-Based[7], т.к. имеют высокое соотношение размера входного сообщения к размеру генерируемого текста, и получающийся текст, максимально похож на естественный. Стоит отметить, что стеготекст, как правило, является бессмысленным.

Существует обратная стеганографии задача – стегоанализ. Его целью является выявление факта наличия встроенного секретного сообщения в контейнере. В качестве критерия оценки эффективности методов стегоанализа используют вероятность обнаружения секретного сообщения в контейнере или вероятность возникновения ошибки. Существует два рода ошибок:

− *ошибка 1-го рода* – случай, когда метод принимает пустой контейнер (без секретного сообщения) за заполненный (с секретным сообщением).

− *ошибка 2-го рода* – случай, когда заполненный контейнер принимается за пустой.

Как уже было отмечено, использование методов, генерирующих текст, подобный естественному, имеет один недостаток − получается бессмысленный текст. Задача определения осмысленности текста требует участия человека. Однако, учитывая большой объем передаваемых сообщений в сети, это не всегда возможно. Поэтому особенно актуальна задача создания эффективных средств компьютерного анализа, работающих без участия человека.

На сегодняшний день, существует большое число различных методов компьютерного стегоанализа. Рассмотрим более подробно следующие методы:

Метод, использующий недостатки семантических методов, опубликованный в работе [8]. При замене слов на их синонимы существует вероятность нарушения правил семантики языка, например: При встраивании сообщения в предложение "What time is it?" слово *time* может быть заменено на *period* или *duration,* что не является корректным для английского языка. При определении текста, полученного программой **Tyrannosaurus Lex,** ошибка 1-го рода составляет 38.6%. Ошибка 2-го рода – 15.1%. Столь большие значения ошибок затрудняют практическое применение метода. Кроме того, данный метод требует достаточно много времени работы, а также большую базу данных правил языка.

Метод, предложенный в работе [9], использует частоту встречаемости слов и ее дисперсию в анализируемом тексте. По полученным данным с помощью SVM классификатора определяется факт наличия стеготекста, сгенерированного программными средствами [6],[7] или [8] в контейнерах размером 5Кб и более. Сумма ошибок 1-го и 2-го рода не превосходят 7.05%.

Метод, предложенный в работе [10], базируется на анализе статистической зависимости слов в тексте. Такая зависимость известна для стеготекста, обычного текста. С помощью специального алгоритма производится сбор статистики подозрительного контейнера и, используя SVM классификатор, определяется наличие стеготекста, сгенерированного программными средствами [6],[7] или [8] в контейнере. Метод имеет ошибку 1-го и 2-го рода в сумме 12.61%, 4.49%, 1.5%, 0.85%, 0.43% на текстовых сегментах размером 5Кб, 10Кб, 20Кб, 30Кб и 40Кб соответственно.

Наиболее эффективным, при малых размерах входных данных является метод, опубликованный в работе [11], строящий модель языка текстов. После чего по имеющимся моделям стеготекста, обычного текста и текста контейнера, используя SVM классификатор, определяется – является ли подозрительный текст обычным или стеготекстом. Точность обнаружения текста, сгенерированного программой [5], составляет 99.61% на текстовых сегментах размером 400 байт и более.

Итак, подытожим результаты наиболее эффективных существующих методов стегоанализа:

Таблица 1. Эффективность существующих методов стегоанализа

| Метод стегоанализа | Атакуемая стегосистема | Размер контейнера (байт) | Вероятность правильного обнаружения стеготекста |
|---|---|---|---|
| [11] | Nicetext | 400 | 99.61% |
| [9] | Texto | 5000 | 92.95% |
| [9] | Marko-Chain-Based | 5000 | 92.95% |

В данной работе предлагается эффективный метод выявления скрытых сообщений, полученных с помощью метода [4], для текстовых сегментов размером в 400 байт, сгенерированных программой Texto. Точность определения стеготекста составляет не менее 99.98%.

## 2. Описание предлагаемого метода

В настоящей статье предлагается новый метод, основанный на подходе, предложенном в работе Рябко Б.Я. [12], отличающийся от других тем, что для выявления факта наличия "стеготекста" используется сжатие обычным архиватором. Идея подхода состоит в том, что внедряемое сообщение нарушает статистическую структуру контейнера, повышая его энтропию. Следовательно, заполненный контейнер будет "сжиматься" хуже, чем незаполненный. В отличие от предыдущих методов, данный метод обладает рядом преимуществ: Анализ занимает сравнительно мало времени (порядка 0.1-0.5 сек на современных персональных компьютерах). Для проведения анализа не требуется словарей синонимов или правил грамматики языка, занимающих большой объем памяти.

Теперь рассмотрим основную идею предлагаемого метода на следующем примере. Пусть существуют контейнеры A и B. Один из них, содержит стеготекст. Размеры контейнеров до и после сжатия архиватором следующие:

Таблица 2. Размеры контейнеров

|   | До сжатия | После сжатия |
|---|---|---|
| A | 500 | 300 |
| B | 500 | 320 |

Возьмем C – контейнер содержащий стеготекст. Допишем C в конец контейнеров A и B, сожмем и сравним получившиеся длины сообщений C до сжатия и после.

Таблица 3. Размер сообщения

|   | До сжатия | После сжатия |
|---|---|---|
| C | 50 | 45 |
| C | 50 | 20 |

Исходя из полученных размеров до и после сжатия можно утверждать, что контейнер C является статистически зависимым относительно контейнера B, что обеспечивает хорошее сжатие. И наоборот, является независимым относительно A, то есть сжатие будет хуже. Следовательно, стеготекстом является контейнер B. На этом принципе строится стегоатака на широко известное программное средство Texto [6].

Теперь перейдем к непосредственному описанию метода. Перед анализом содержимое контейнера должно быть обработано специальным образом:
− символы, не относящиеся к цифрам, латинским буквам, пробелу, переводу строки, и знакам препинания должны быть удалены;
− последовательности из двух или более символов пробела или перевода строки заменяется на один символ соответственно;
− подготовленный контейнер должен иметь размер в 400 байт (лишнее следует удалить).

Допишем в специально подобранные тексты (полученные с помощью программ [5] и [6]) размером $N$ и $T$, содержимое анализируемого контейнера. Обозначим получившиеся размеры текстов $N_x$ и $T_x$ соответственно. Далее производится сжатие имеющихся текстов с помощью архиватора Bzip2 (использует преобразование Барроуза-Уилера). Затем производится расчёт статистических характеристик по следующим формулам: (индекс *bzip2* - означает размер после сжатия):

$$\alpha = \left(\frac{N_{bzip2}}{N} - \frac{Nx_{bzip2}}{Nx}\right)*100\% \quad \beta = \left(\frac{T_{bzip2}}{T} - \frac{Tx_{bzip2}}{Tx}\right)*100\%$$

Определение факта наличия или отсутствия стеготекста в контейнере осуществляется исходя из полученных характеристик ($\alpha$ и $\beta$). Эмпирическим путём в ходе ряда экспериментов были подобраны оптимальные значения ($\alpha$ и $\beta$). Характеристики обычного текста должны удовлетворять следующему условию: *($\alpha>0.9$)ИЛИ($\beta<1$)*. Если условие не выполняется, то это означает наличие стеготекста в контейнере.

## 3. Экспериментальный анализ

Для сравнения нового метода с ранее известными необходимо определить эффективность метода экспериментально. В качестве критерия эффективности возьмём процентное отношение количества правильно определённых фактов наличия или отсутствия стеготекста в контейнере к общему числу попыток определения.

Для эксперимента была сформирована выборка, состоящая из 10000 случайно отобранных файлов содержащих обычный текст и 10000 содержащих стеготекст. Каждый файл, содержащий стеготекст, был получен с помощью программы Texto. Для этого вход программе подавался файл, содержащий псевдослучайную последовательность (имитирующий зашифрованное сообщение). На выходе программы создаётся файл содержащий стеготекст. Для каждого файла выборки производился анализ с помощью разработанного программного средства, рассчитывались характеристики, по которым определялось наличие стеготекста. Были получены следующие результаты:

Таблица 4. Результаты работы программы

| Содержимое контейнера | Количество правильных определений наличия/отсутствия стеготекста | Количество неправильных определений |
|---|---|---|
| Обычный текст | 10000 | 0 |
| Стеготекст | 10000 | 2 |

По результатам анализа построен график зависимости точности определения (в процентах) наличия или отсутствия стеготекста от размера контейнера (в байтах)

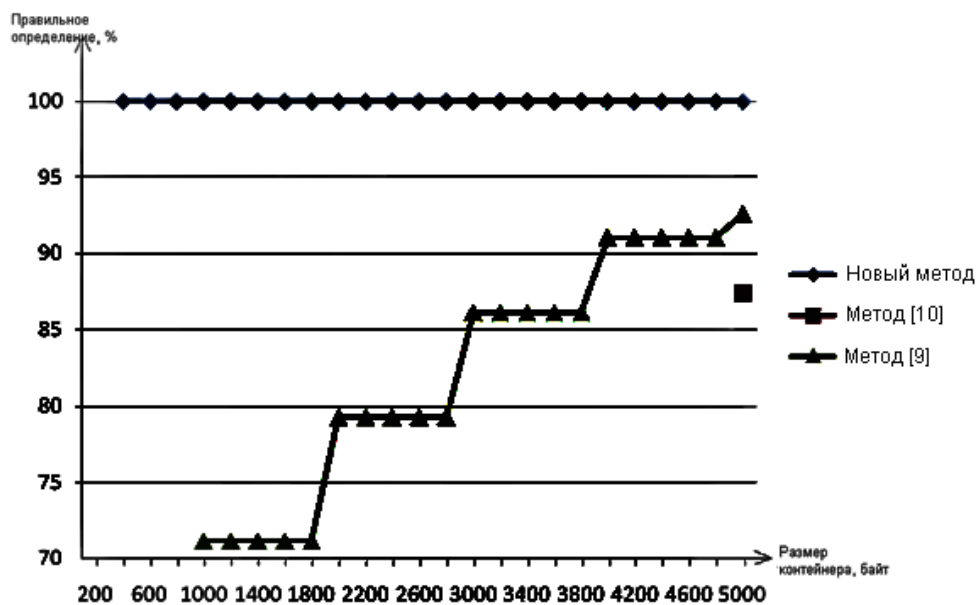

Рисунок 1. График эффективности методов стегоанализа

Таким образом, мы видим, что предложенный метод позволяет эффективно решать задачу определения скрытой информации. Точность предлагаемого метода составляет не менее 99,98% для текстовых сегментов размером 400 байт, полученных программой Texto[1].

Подведем итог результатов работы наиболее эффективных методов стегоанализа:

Таблица 5. Эффективность методов стегоанализа

| Метод стегоанализа | Атакуемая стегосистема | Размер контейнера (байт) | Вероятность правильного обнаружения стеготекста |
|---|---|---|---|
| [11] | Nicetext | 400 | 99.61% |
| Новый метод | Texto | 400 | 99.98% |
| [9] | Marko-Chain-Based | 5000 | 92.95% |

Очевидно, что данный метод превосходит по эффективности, другие современные методы стегоанализа.

---

[1] Результат был независимо проверен М. Ю. Жилкиным

**Нечта Иван Васильевич**
аспирант, ассистент кафедры прикладной математики и кибернетики СибГУТИ, тел. (383)2-698-272, e-mail: **www@inbox.ru**